\documentclass[acmsmall, anonymous=False]{acmart}

\renewcommand\footnotetextcopyrightpermission[1]{}
\copyrightyear{}
\acmYear{}
\acmDOI{}
\acmConference[]{}{}{}
\acmBooktitle{}
\acmPrice{}
\acmISBN{}

\usepackage{makecell}
\newcommand{\lorenzo}[1]{\textcolor{black}{#1}}

\begin{document}

\title[Perceptions of Diversity in Electronic Music]{Perceptions of Diversity in Electronic Music: the Impact of Listener, Artist, and Track Characteristics}
\author{Lorenzo Porcaro}
\email{lorenzo.porcaro@upf.edu}
\affiliation{%
  \institution{Music Technology Group, Universitat Pompeu Fabra}
  \city{Barcelona}
  \country{Spain}
}

\author{Emilia G\'{o}mez}
\email{emilia.gomez@upf.edu}
\affiliation{%
  \institution{Music Technology Group, Universitat Pompeu Fabra}
  \city{Barcelona}
  \country{Spain}
}
\affiliation{%
  \institution{Joint Research Centre, European Commission}
  \city{Sevilla}
  \country{Spain}
}

\author{Carlos Castillo}
\email{carlos.castillo@upf.edu}
\affiliation{%
  \institution{Web Science and Social Computing Group, Universitat Pompeu Fabra}
  \city{Barcelona}
  \country{Spain}
}

\renewcommand{\shortauthors}{Porcaro, G\'{o}mez and Castillo}

\begin{abstract}
Shared practices to assess the diversity of retrieval system results are still debated in the Information Retrieval community, partly because of the challenges of determining what diversity means in specific scenarios, and of understanding how diversity is perceived by end-users.
The field of Music Information Retrieval is not exempt from this issue.
Even if fields such as Musicology or Sociology of Music have a long tradition in questioning the representation and the impact of diversity in cultural environments, such knowledge has not been yet embedded into the design and development of music technologies.
In this paper, focusing on electronic music, we investigate the characteristics of listeners, artists, and tracks that are influential in the perception of diversity.
Specifically, we center our attention on
1) understanding the relationship between perceived diversity and computational methods to measure diversity, and
2) analyzing how listeners’ domain knowledge and familiarity influence such perceived diversity.
To accomplish this, we design a user-study wherein listeners are asked to compare pairs of lists of tracks and artists, and to select the most diverse list from each pair.
We compare participants’ ratings with results obtained through computational models built using audio tracks’ features and artist attributes.
We find that such models are generally aligned with participants’ choices when most of them agree that one list is more diverse than the other.
In addition, we observe how differences in domain knowledge, familiarity, and demographics influence the level of agreement among listeners, and between listeners and computational diversity metrics.

\end{abstract}



\maketitle

\section{Introduction}
\label{introduction}
Research on diversity in Information Retrieval (IR) typically seeks diversity as a multiple criteria objective, or criteria based on multiple diversity axioms.
A diverse set of items should be, among other things, a list in which each item is \textit{novel} with respect to those previously seen, in which items cover different potential \textit{intents} of the user, and convey different \textit{aspects} of the information sought \cite{amigo18}.
Computational approaches, adopted in a number of IR contexts, enable the estimation of measurements that can serve as proxies of these criteria within specific tasks, such as subset selection \cite{Drosou2017,mitchell2020}, ranking \cite{Clarke2008,Gollapudi2009}, and recommendation \cite{Kunaver2017}.
A complementary approach to diversity seeks to understand and ideally model what users \textit{perceive} as diversity, a challenge that is addressed through user studies and fraught with difficulties \cite{Sakai2019}. 
Several works confirm that characteristics such as demographics \cite{Park2016}, personality traits \cite{chen2013}, or domain knowledge and expertise \cite{Knijnenburg2012} influence individuals’ diversity needs and perceptions.
In general, what a system considers to be \textit{diverse} may be perceived differently by people interacting with it.

The present study investigates the interrelationship between diversity aspects related to listeners, artists, and tracks in the music domain.
These actors have been represented as the threefold nature of the music phenomenon \cite{Molino1990}: “\textit{as an arbitrarily isolated object} [a track], \textit{as something produced} [by an artist], \textit{and as something perceived} [by a listener]”. 
Our research aims to deepen our understanding of how different diversity facets interplay in the music field and how are they experienced:  “[..] \textit{subjectivity as coloured by embodied experience, which in human societies is mediated by such social differences as gender, class, race and ethnicity – so that subjectivities, experience and embodied experience are not everywhere the same}.”\cite{Born2020}.

We designed a user study to investigate diversity in the music domain, specifically in the Electronic Music (EM) genre. 
The study was divided into three parts. 
In the first one, the stimulus consisted of two sets of four tracks each. 
We asked participants to briefly listen to the tracks and evaluate which, among the two sets, they perceived to be more diverse. 
In the second part, the stimulus was two sets of photos of four artists each, along with some basic information about them. 
We asked participants to indicate which set was more diverse according to socially salient attributes (e.g., age and gender). 
In the following, we refer to part one as the \textit{TRACK} task and part two as the \textit{ARTIST} task. 
A third part which we call \textit{COMBO} combined the \textit{TRACK} and \textit{ARTIST} tasks.

\begin{figure}[h]
\centering
\includegraphics[width=0.85\textwidth]{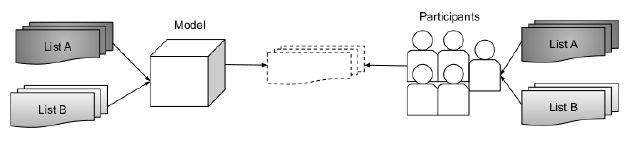}
\Description{High-level view of the study. On the left side, two lists are provided to a computational model designed to identify the most diverse one, while on the right side the same lists are provided to a group of people. By analyzing the agreement among the two processes, we aim at highlighting the key factors in diversity estimations.}
\caption{High-level view of the study. On the left side, two lists are provided to a computational model designed to identify the most diverse one, while on the right side the same lists are provided to a group of people. By analyzing the agreement among the two processes, we aim at highlighting the key factors in diversity estimations.}
\label{fig:fig1}
\end{figure}

We built a computational model to automatically estimate diversity by considering distance measures between tracks based on specific music-related features extracted from the audio signal, as well as distance measures between artists based on their attributes. 
Then, we compared participants’ ratings to this model, performing inter-rater reliability analysis (IRR) and metric reliability analysis (MR). 
Additionally, we compared the results from all participants against subgroups created by clustering participants based on their EM domain knowledge, familiarity, and/or demographics. Figure \ref{fig:fig1} depicts the high-level structure of the experiment. 

The purpose of this analysis is to answer the following research questions:
\begin{enumerate}
    \item \textbf{RQ1. To what extent tracks’ audio features and artists’ attributes can be used to assess perceived diversity?}
    \item \textbf{RQ2. To what extent domain knowledge and familiarity influence participants’ perceptions of diversity?}

\end{enumerate}

Our main finding regarding RQ1 is that computational methods can to some extent be used to compare diversity between two lists, especially when there is a strong agreement among listeners that one list is more diverse than the other. 
Our main findings regarding RQ2 are that domain knowledge and familiarity with a music genre are partially aligned, and that in general listeners with more domain knowledge agree less with automatically extracted metrics, and disagree more with each other.

We develop these and other findings in the rest of the paper as follows: Section \ref{background} provides an overview of how diversity has been approached from an information processing perspective, and surveys diversity and agreement studies performed in the Music Information Retrieval field. 
Following, Section \ref{survey} reports the method used in this work, describing the user study design, while Section \ref{features} the considered features and metrics. 
Section \ref{results} presents the results of our analysis, discussed together with their limitations in Section \ref{discussion}. 
Conclusions are presented in Section \ref{conclusion}.

\section{Background and Related Work}
\label{background}
Next, we introduce the concept of diversity for then reviewing its related literature in ranking and recommendation. 
Afterwards, focusing on the \lorenzo{Music Information Retrieval (MIR)} literature we first present diversity-related user studies on music listening, and we then review the work done in the analysis of annotators' agreement.

\subsection{The Value of Diversity}
The definition and operationalization of diversity have been the object of study of scholars from several disciplines through history, ranging from ecology to information science, economic and cultural studies \cite{stirling2007}.
Such a plethora of approaches investigated several concepts of what can be intended as \textit{diversity}, which according to the context can be defined and measured differently, with different implications as discussed in detail by Steel et al. in \cite{Steel2018}. 
Nonetheless, what generally is shared among academic communities, but also public and private institutions, is the importance of fostering diversity as a tool to enhance creativity but also productivity \cite{stirling2007}, to promote pluralism and equality \cite{unesco2001}, to make people aware of different viewpoints and facilitate the public debate \cite{Helberger2011}.

Even if these benefits have been proven in several areas, \lorenzo{simultaneously} major concerns have emerged due to the lack of diversity, and Artificial Intelligence (AI) is a notable example of a field in which we are witnessing a diversity crisis \cite{West2019, Freire2021}.
Indeed, AI systems --- an umbrella term wherein also Recommender Systems can be included \cite{Jannach2020} --- have already been proven to reinforce hegemonic biases and discrimination in different applications, in fields such as Computer Vision \cite{Raji2020} and Natural Language Processing \cite{Bender2021}.
To contrast these harmful effects, in the process of design, development, implementation, and evaluation of these systems, diversity has been already identified as a key requirement to build trustworthy AI systems \cite{AIHLEG2019}.

In the music field, scholars from music psychology, sociology, and ethnomusicology questioned under different perspectives the value of diversity \cite{Huron2004, Grenier1989, Clarke2015}, stressing the importance of pushing the boundaries of researching and learning beyond Western music traditions. 
On the contrary, few examples in the MIR literature can be found where multicultural analysis has been carried out, e.g. \cite{Serra2011}, and the lack of diversity in this field has been already acknowledged in different contexts \cite{Born2020}. 
However, recently the effort of the MIR community in promoting initiatives like Women in MIR\footnote{\url{https://wimir.wordpress.com}}, and also organizing within the International Society for Music Information Retrieval (ISMIR) conference, the main forum of MIR practitioners, a special call for Cultural Diversity papers\footnote{\url{https://ismir2021.ismir.net/cfp}}, make evident how nowadays diversity has been acknowledged as core topic on which more research and awareness is needed.

\subsection{Diversity in Ranking and Recommendation}

Information Retrieval systems have benefited from the introduction of evaluation practices in which diversity is key to determine whether results are useful for end-users. 
In ranking scenarios, the Probability Ranking Principle (PRP), affirming that documents should be ranked by their probability of being relevant in decreasing order, has been shown to be unable to deal with concepts such as ambiguity and redundancy \cite{Chen2006}. 
In the well-known example in the IR community of the ambiguous query “Jaguar”, a retrieval system for satisfying the possible information needs of a user is expected to return documents related to, e.g., an animal, a carmaker, and a brand of guitars, and diversity may aid in expanding the PRP offering a new perspective from which to rank relevant documents \cite{Clarke2008}. 
Diversification techniques recognize that the relevance of a document depends not only on its nature but also on its relationships with other documents \cite{Agrawal2009}. These techniques are often based on axiomatic approaches describing the properties that diversification systems and diversity metrics should satisfy \cite{Gollapudi2009, amigo18}.

Parallelisms between ad-hoc IR and Recommender System (RS) diversification techniques have naturally emerged, given that the recommendation task can be seen as an IR task \cite{Vargas2011}. 
Similarly to what happens in ranking settings, redundancy in RS can create a so-called \textit{portfolio effect}, i.e., recommendations provided to users are too similar to each other \cite{Burke2002}. 
As a consequence, RS research has highlighted the relevance of going beyond accuracy metrics when evaluating the retrieved results to increase the overall users’ satisfaction \cite{Ziegler2005, McNee2006}. 

Centering on user studies, Sanderson et al. \cite{Sanderson2010} compared different retrieval systems and found that users tend to prefer systems that, \lorenzo{according to the considered diversity metrics,} return \lorenzo{more} diverse outcomes. 
A similar result is presented in \cite{castagnos2013}, where authors showed that users are more satisfied with recommendations provided by \lorenzo{the system optimized to maximise the proposed diversity metric}. 
Such results are confirmed in \cite{Willemsen2016}, which proves that diversification techniques help in reducing users' choice difficulty and choice overload. 
\lorenzo{Moreover}, in \cite{Ekstrand2014,Zhao2017} authors found that users' satisfaction is positively correlated with the diversity of the provided recommendations.

Nonetheless, even if in the last decades the IR and RS communities have actively researched diversity, there is still an ongoing debate around whether such diversity metrics are effectively beneficial from a user point of view \cite{Sakai2019}. 
Generally, offline evaluation can be easily performed to understand points of strength and weakness of retrieval systems, while user studies are more difficult to execute and rarer in the literature \cite{Hearst1999, Pu2012}. 
The interactions between humans and retrieval systems are particularly challenging to analyze due to the differences among users' characteristics and needs, and to collect such data is a time-consuming task and expensive in terms of resources.

\subsection{Diversity in Music Listening User Studies}
Diversity has been empirically investigated by the MIR community in relationship with listening behaviours, mostly focusing on its application in recommendation scenarios. 
In \cite{Ferwerda2016}, Ferwerda et al. grouped listeners by their country of origin, and then analyzed how different diversity needs are linked to each country's cultural dimensions \cite{hofstede2005cultures}. 
Farrahi et al. \cite{Farrahi2014} showed how recommender systems performance increases for users with greater diversity in terms of listened music genres. 
Furthermore, analyzing the diversity perceived by listeners when exposed to a recommendation list, a positive correlation has been found between the perceived diversity, recommendation attractiveness, and the possibility to discover new music \cite{Ferwerda2017}. 
The same authors also explored the link between musical sophistication and diversity needs, showing how a higher sophistication corresponds to more diversified listening behaviours \cite{Ferwerda2019}. 
In addition, through semi-structured interviews, authors in \cite{Robinson2020} examined what listeners look for when faced with diversity-aware music recommendations, characterizing participants’ inner and outer diversity as diversity within and outside their own music preferences. 
In most of these studies, users’ listening behaviours are characterized by their diversity, computed using as a proxy the number of different artists, tracks, or music genres listened to by users. 
In contrast with previous work, we focus on the exploration of what are the characteristics that are associated with perceived diversity in the music domain.

\subsection{Human Annotations Agreement in Music Information Retrieval}
Our study features heavily human annotations over music, which is why we present here some background on this topic, connecting to previous works from the MIR literature.

The analysis of the agreement between human annotators is considered, among other factors, to influence the reliability of scientific experiments, giving a more robust interpretation of the obtained results \cite{Urbano2013}. 
While in the IR field evaluation practices have a long tradition, and are a subfield of research on their own, they are comparatively not as well established in MIR and most of the work done has focused on the concept of music similarity, more than diversity \cite{Schedl2014}. 
Indeed, because of its relevance in several MIR-related tasks, such as music classification, cover detection, and music recommendation, the MIR community has posed great attention in developing proper evaluation practices for music similarity \cite{Urbano2012}. 
Flexer et al. \cite{Flexer2014, Flexer2016} showed that due to the highly subjective, context-dependent, and multi-dimensional nature of music similarity, the lack of inter-rater agreement between annotators provides an upper bound of the performance of retrieval systems based on such notion. 
A recent study showed how selecting a controlled group of subjects and music material can lead to an increase in inter-rater agreement, but still questioning the validity of experiments on general music similarity \cite{Flexer2019}.

From another perspective, the inter-rater agreement has been analyzed in the context of Music Emotion Recognition (MER), another task in which the high subjectivity in estimating emotion perceived when listening to music has limited the applicability of algorithmic solutions. 
Schedl et al. \cite{Schedl2018} analyzed the relationship between listeners’ characteristics and emotion perceived while listening to classical music, founding a greater agreement for non-expert classical music listeners. 
Similarly, in \cite{Gomez-Canon2020} authors investigated the role of personal characteristics when classifying annotations related to music, focusing on the rock and pop music genres, showing that annotators have a higher agreement for basic emotions than for complex ones. 
A further example of annotators' agreement analysis is presented in \cite{Oramas2016}, where agreement is used to evaluate the performance of information extraction methods for building a knowledge base in the music domain.

Following previous research, this study analyzes annotators' agreement to understand how diversity is perceived among different groups of listeners, studying also the agreement between them and diversity metrics as discussed in the following sections. 
On one hand, measuring metric reliability we seek agreement as confirmation of the proposed diversity metrics. 
On the other hand, using inter-rater reliability we evaluate if data are interpreted in the same way among the participants.
This is in line with norms and guidelines for CSCW and HCI practice \cite{McDonald2019}.

\section{Study Design}
\label{survey}

We designed an online survey for our user study. A facsimile of the survey, along with anonymized responses, is available with our data release.\footnote{\lorenzo{\url{https://zenodo.org/record/4436649}}} The study was an activity of the \lorenzo{TROMPA (Towards Richer Online Music Public-domain Archives)  project\footnote{\lorenzo{\url{https://trompamusic.eu}}}} and followed the personal data management and ethical protocols approved within this project.

\subsection{Recruitment and Informed Consent}
\label{recruitment}
We recruited participants by sharing a landing page from where to access the survey, using the authors’ personal and institutional social networks (Twitter, Facebook, Reddit, and Instagram), and MIR-related mailing lists.
Responses were collected from July through October 2020.

Participants were shown details of the survey and an information sheet describing the research objectives, methodology, risks, and benefits.
They were informed that participation was voluntary and could be withdrawn at any point, and of their rights including the right to access, rectify, and delete their information.
Personal data collection was limited to what the participant chose to disclose, as all demographic and background questions were labeled as optional.
At the end of the survey, participants could opt-in to be asked further questions, and/or to receive the results of their study by e-mail; in that case, they could provide their email address.
This part was also clearly marked as optional.

\subsection{Survey Sections}
\label{sections}
\begin{figure}[h]
\centering
\includegraphics[width=0.9\textwidth]{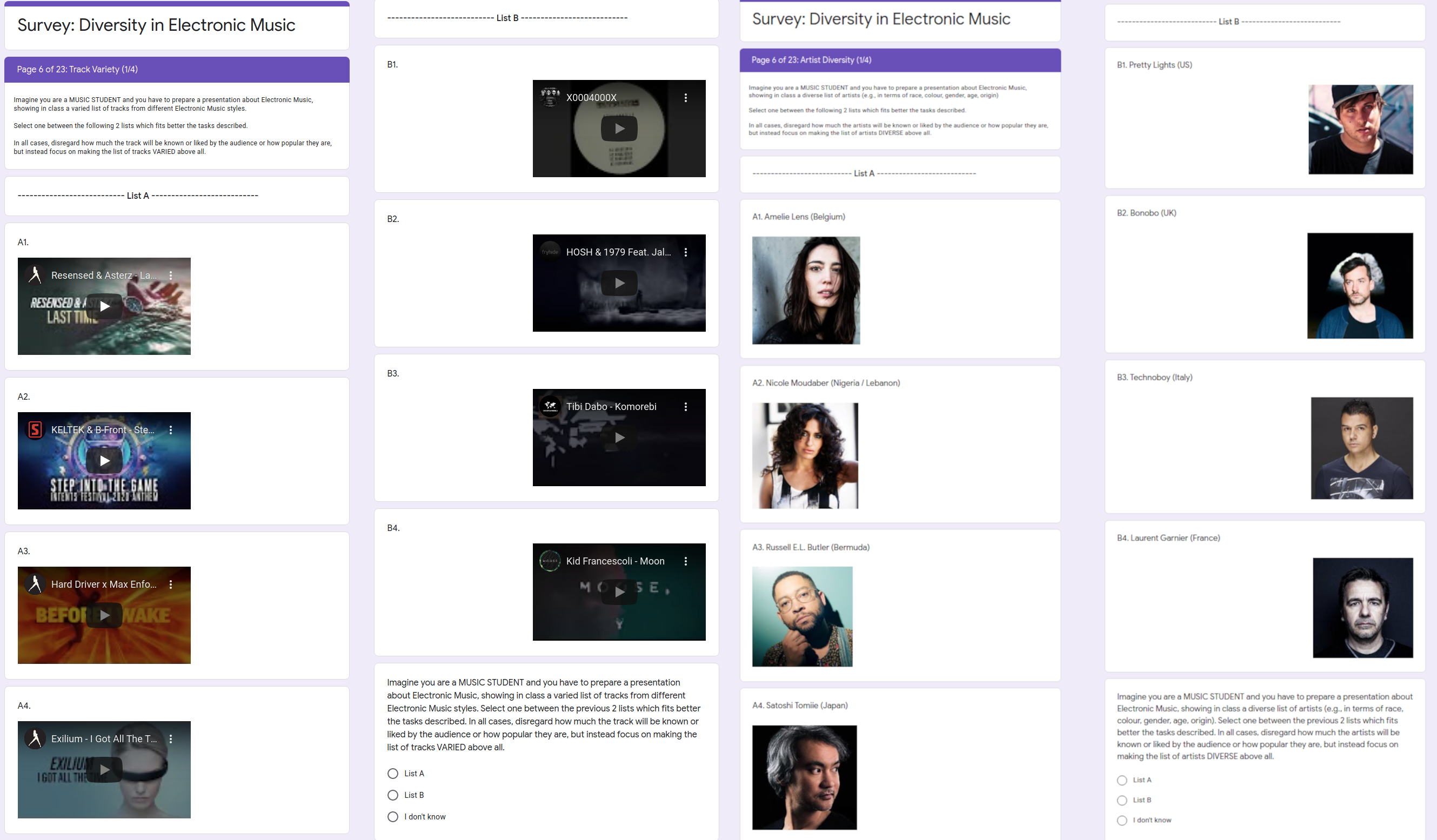}
\Description{Screenshot of the interface for the \textit{TRACK} (left) and \textit{ARTIST} (right) tasks.}
\caption{Screenshot of the interface for the \textit{TRACK} (left) and \textit{ARTIST} (right) tasks.}
\label{fig:fig2}
\end{figure}

The survey was composed of five main sections:
\begin{enumerate}
    \item \textit{Participants’ information collection}.
    Optional demographic information, with each question prominently labeled as optional: age, gender, skin tone according to the Fitzpatrick scale \cite{Fitzpatrick1988}, continent origin, level of education. \lorenzo{With regards to the choice of including skin tone, we agree with the work by Mitchell et al. \cite{mitchell2020} that while introducing a diversity metric it is critical to consider attributes "[...]\textit{defined in light of human attributes involved in social power differentials, such as gender, race, color, or creed}". Under this lens, we consider skin tone as one of the social salient characteristics which may impact the perception of diversity. Nonetheless, we acknowledge that Fitzpatrick scale is an approach to skin classification that is not exempt from racial limitations, as discussed in \cite{Ware2020}.}
    As detailed in Section \ref{statistics}, the \lorenzo{demographic} information has been collected in order to 1) describe the population of the experiment, and 2) cluster participants in subgroups based on demographic characteristics.
    \item \textit{Domain knowledge}.
    Optional information about participant’s music background and familiarity with electronic music: whether they had any formal music training (yes/no), music playing engagement, taste variety, EM listening frequency, EM taste variety (5-Point Likert scale).
    This was accompanied by a test for estimating participants’ familiarity with EM artists (detailed in Section \ref{domainknowledge}).
    \item \textit{Track Variety (\textit{TRACK}) task}.
    Participants were asked to listen to two sets of audio tracks and to choose the most diverse in terms of music features.
    Tracks were shown by means of YouTube embeddings, without specifying additional information such as tracks’ title or artist (see Figure \ref{fig:fig2}).
    Participants were explicitly asked to not consider the tracks’ popularity, their preferences, or their familiarity with the tracks, but to focus only on selecting the most diverse list within each pair.
    They were also not asked to listen entirely to every track, but to navigate through them as they normally would do when listening to music in a digital environment.
    \item \textit{Artist Diversity (\textit{ARTIST}) task}.
    Two lists of artists were shown to participants, then asked to choose the most diverse one in terms of socially relevant attributes.
    Artists were presented using close-up or half-length photos, avoiding when possible images where they are performing or acting (see Figure \ref{fig:fig2}).
    Together with the photo, the name of the artist and the birthplace were shown.
    Again, participants were explicitly asked to not consider artists’ popularity, their preferences, or their familiarity with the artists when selecting the most diverse list in each pair.
    \item \textit{Combined (\textit{COMBO}) task}.
    Two lists of tracks and corresponding artists were shown to participants, asked to indicate the set most diverse in terms of tracks’ music features and artists’ attributes at the same time.
    The information displayed for every list was the sum of the previous tasks.
    The indications for selecting the list given to the participants were the same: to not consider popularity, personal preferences, or familiarity when choosing the most diverse list.
\end{enumerate}

While the \textit{TRACK} task is the most realistic of all and corresponds more specifically to how people consume music, \textit{ARTIST} and \textit{COMBO} tasks instead can be considered experimental artifacts to complement the observations from the first. \lorenzo{Several works in multimodal Music RS research, in addition to audio signals, exploit the so-called contextual information such as album covers \cite{Lin2018}, or artist biographies \cite{Oramas2017}. This is motivated by the fact that the content itself (i.e. the audio in the music domain) does not carry all the music information, and therefore methods including other types of data have recently emerged, also thanks to the advancement of Deep Learning techniques. Similarly, we believe that to assess the diversity of a music list, the audio may be the component with the most influence on the perception but not the only one.}

In each task (\textit{TRACK}, \textit{ARTIST}, and \textit{COMBO}), participants had to compare four pairs of lists, each list formed by four elements (i.e., tracks, artists, and tracks and artists together).
For each comparison, they had to select between the first list (“\textit{List A}”), the second one (“\textit{List B}”), or neither (“\textit{I don’t know}”).
At the end of each task, participants were asked to provide feedback about their selection strategy, indicating the factors that most influenced their choices.
To avoid order effect bias, we created four versions of the survey, switching the order of the \textit{TRACK} and \textit{ARTIST} tasks (parts 3 and 4 above), and the order of the lists within each pair.
The survey was implemented using Google Forms and available in English, Spanish, and Italian: participants were able to select the preferred language for completing the survey on the landing page.

\subsection{Domain Knowledge and Familiarity}
\label{domainknowledge}
We have been able to estimate participants' familiarity with EM artists, independently from the previously self-declared answers, by designing a quiz presented at the end of the first part of the survey, in which they were asked to select which among a list of \textit{n=30} selected artists they know, they do not, or they maybe know.
The list was composed of 15 artists picked within the EM highlights of AllMusic,\footnote{\url{ https://www.allmusic.com/genre/electronic-ma0000002572/artists}} a selection curated by music experts and including EM most representative artists, and 15 artists selected within the ones shown in the survey tasks, mostly less-known and not mainstream.
Artists were presented in a randomized fashion to avoid order effect bias.
As a result, we assigned to every participant a familiarity score, computed as an inverse-popularity weighted sum; this reflects the intuition that knowing several of the less popular artists implies more familiarity with EM.

Specifically, the familiarity score was computed with the following procedure.
First, we gather several social signals for every artist on the list: Spotify, Twitter, Facebook, Deezer, and SoundCloud followers, together with Spotify monthly listeners, and Facebook likes.
We aggregate such popularity-like data using the geometric aggregation of popularity (GAP0) metric presented in \cite{Koutlis2020}, a non-linear method for popularity metric aggregation based on geometrical shapes.
The list of artists and the corresponding GAP0 score is presented in Figure \ref{fig:fig3} (see Appendix \ref{appendixA}).
Afterward, we rank artists by their score, and we assign an alternative score inversely proportional to their GAP0 popularity, ranging from $1/n$ for the most popular artist to 1 for the less popular.
The final familiarity score assigned to every participant is a weighted sum of the artists’ alternative score, where participants got: a) the entire score if the artist is known b) half of the score if the artist is maybe known, and c) zero if the artist is unknown.
Finally, participants’ familiarity score is normalized between 0 and 1.

\subsection{Task Selection}
\label{taskselection}
Tracks and artists data have been manually selected by searching in knowledge bases such as \textit{Wikipedia}, \textit{AllMusic}, \textit{MusicBrainz}, \textit{Discogs}, and \textit{Resident Advisor}, but also from artists’ --- websites, social media, and interviews in music magazines. The complete list of tracks and artists used for the \textit{TRACK}, \textit{ARTIST}, and \textit{COMBO} tasks are reported in Appendix \ref{appendixA} (respectively Table \ref{tab:tracktask}, Table \ref{tab:artisttask}, and Table \ref{tab:bothtask}).
The displayed items have been chosen to guarantee that the differences between lists were more pronounced in the former two out of four pairs, while more subtle in the latter two pairs.
In the case of the \textit{TRACK} task, such differences were based on tracks’ audio and semantic features (see Section \ref{trackfeatures}), e.g. showing a list where tracks have different tempo or are associated with various music genres, whilst in the case of the \textit{ARTIST} task, diverse lists are where artists belong to different social groups.
This strategy has been chosen for helping the participants become familiar with the task.
Similarly, the \textit{COMBO} task has been shown after the previous two tasks because of its higher complexity, since the comparison had to be based on both tracks’ and artists’ information.

Only solo artists have been included in the experiment. Indeed, even if they do not represent the totality of the EM scene, we avoid including duos or groups to facilitate the comparison between lists.
Regarding the content included in the survey, while some of the tracks and artists are shared among different tasks, they have been selected independently in each case for being able later to analyze and discuss results separately, as presented in Section \ref{results}.
As consequence, the lack of a major overlap of the lists among tasks made it not possible to analyze robustly how consistent were the participants' judgments across the three tasks, a factor excluded in the initial design of the experiment. 
The evaluation of the diversity using of the distance metrics, later introduced in Section \ref{distancemetrics}, confirms our design strategy as discussed in Section \ref{agreement}.

\section{Features and Measurements}
\label{features}

This section describes the metrics that we computed to estimate the diversity of lists of tracks and artists.

\subsection{Track Features}
\label{trackfeatures}
Two groups of features for computing distances between tracks have been selected from standard features proposed in the MIR literature \cite{Schedl2014}: low-level, rhythmic, and tonal features (LL), and high-level features (HL).
LL features are directly related to audio signal properties, while HL features represent aspects that are more understandable by human perception.
Examples of features included in the LL group are BPM (Beats Per Minute), related to the tempo of the track, or HPCP (Harmonic Pitch Class Profile), describing the pitch and tonal content of the piece.
In the set of HL features, we consider mood-related features, such as “happy/sad” or “aggressive/relaxed”, features describing the presence of voice or instrumental parts, or the darkness/brightness of a track. The complete list of features is included in Appendix \ref{appendixA} (see Table \ref{tab:features}).
Specifically, we used \textit{Essentia} \cite{Bogdanov2013}, an open-source library for audio and music analysis to extract the features from the audio signals. A detailed explanation of the features and of the algorithms used to extract them is publicly available.\footnote{\url{https://essentia.upf.edu/streaming_extractor_music.html}}
Even if LL and HL features represent different characteristics of an audio track, there are some correlations among them.
For instance, in the set of tracks used in the \textit{TRACK} task, the HL features \textit{danceability} and \textit{party} are positively correlated with BPM (Pearson's $\rho=.41$, $p<.05$ and $\rho=.50$, $p<.005$), while \textit{relax} is negatively correlated with it ($\rho=-.28$, $p=.12$).
Similarly, dissonance and spectral entropy are positively correlated with \textit{aggressivity} ($\rho=.38$, $p<.05$ and $\rho=.37$, $p<.05$).

\subsection{Artist Attributes}
\label{artistaattributes}
We focused on four attributes for estimating the diversity of an artists’ set: gender, origin, skin tone, and debut year.
Gender and skin tone are two salient characteristics included in the image set perception study proposed in \cite{mitchell2020}, and artists' birthplace was explicitly presented to the participants during the survey (see Section \ref{survey}). 
Besides, we included the debut year as a characteristic representing part of the cultural aspects of the artists' music background.
In the case of gender, we grouped artists into two groups: male and not-male, including female, non-binary, transgender, and any other gender different than male.
The reason for this grouping was to compare a privileged gender within the music industry (men) against the other genders \cite{Eppsdarling2020}.
For the origin, we used as a proxy the artists’ birthplace and then built three classifications: at a country-level, continent-level, and using regional groupings proposed by the United Nations in the context of the Millenium Development Goals.\footnote{\url{http://mdgs.un.org/unsd/mdg/Host.aspx?Content=Data/RegionalGroupings.htm}}
For the skin tone, we first classified artists using the Fitzpatrick Scale, and then grouped them in white (skin type I and II) and not-white (skin type III, IV, V, and VI).
Lastly, using the debut year, which represents a music aspect (a moment in electronic music history) but also a generational difference among artists, we divided artists into four groups: 1980s, 1990s, 2000s, 2010s.
%
%
For every attribute, artist groups were balanced considering socio-political power differentials: male versus not-male, white versus non-white, developed regions versus developing regions, and recent versus non-recent artists.

\subsection{Distance Metrics}
\label{distancemetrics}
The measurement of track and artist diversity was accomplished using two distinct metrics, respectively the cosine distance \cite{Singhal01}, and the Goodall distance \cite{Boriah2008}.
This choice is motivated by the difference between the considered tracks' features and artists' attributes.
Indeed, in the former case we estimated the distance using as input a vector composed of normalized numerical values, while in the latter we used categorical data.
Cosine distance has the advantage of being 0 for identical vectors and 1 for orthogonal ones.
The Goodall metric has the advantage of assigning a lower distance to a match when the attribute value is rare.
For instance, in a set formed by ten artists, seven from Northern America and three from Latin America and the Caribbean, all other things equal, two artists coming from Northern America will score a higher distance in comparison to two artists coming from Latin America and the Caribbean, because of the larger presence in the set of the former in comparison to the latter.

The assessment of the most diverse list among each pair was accomplished using a two-step procedure:
\begin{enumerate}
    \item the pairwise distance between items in a list was computed ($d_i$), using the cosine metric for the tracks’ features and the Goodall metric for the artists’ attribute.
    \item using two aggregation operations, namely the minimum pairwise distance between items, $\operatorname{min}(d_i)$, and the average pairwise distance between items, $\operatorname{avg}(d_i)$, we identified the most diverse list as the one with a higher value obtained after aggregating.
\end{enumerate}
Such strategies were chosen following the work by Mitchell et al. on diversity and inclusion metrics for subset selection \cite{mitchell2020}.
Using the $\operatorname{min}(d_i)$ mechanism, the most diverse list is assured to have all the items more diverse among each other than the other list, while in the case of $\operatorname{avg}(d_i)$ the list with a higher diversity has on average items more diverse between each other.
As observed in \cite{mitchell2020}, these two mechanisms can be mapped to two approaches from social choice theory, namely \textit{egalitarian} \cite{sep-egalitarianism} and \textit{utilitarian} \cite{sep-utilitarianism-history}.

\section{Results}
\label{results}

This section presents the results of our survey, which are summarized and discussed in Section~\ref{discussion}.

\subsection{Participant’s Statistics}
\label{statistics}

After filtering out participants less than 18 years old, and participants who did not express any preference in any task (i.e., selecting always “\textit{I don’t know}”), we obtained a total of 115 answers.

Most of the participants are aged between 18 and 35 (61\%), male (73\%), come from Europe or North America (73\%), have a bachelor’s degree or higher (85\%), and identified themselves to have a skin tone type I (pale white skin) or II (white skin) according to the Fitzpatrick scale (60\%).
In the light of these numbers, we can affirm that the population of our study is formed for the most part by subjects coming from WEIRD (Western, Educated, Industrialized, Rich, and Democratic) societies \cite{Henrich2010}.
Regarding their music background and expertise, more than half of the participants indicated that they have received formal music training beyond the usual music lessons in school (58\%). 
Moreover, using a 5-point Likert scale they declared on average to moderately engage in playing, DJ-ing, or producing any kind of music (M=3, IQR=4), to have high to moderately varied listening habits (M=4, IQR=1), to be EM listeners (M=5, IQR=1), and within EM to listen to different styles (M=4, IQR=2).

The familiarity test with EM artists (see Section \ref{domainknowledge}) gave us another perspective to analyze participants.
On average, they obtained a score of 0.3 (on a scale from 0.0 to 1.0) with a standard deviation of 0.2.
Looking at the score distribution grouping participants by their self-declared frequency of EM listening, we notice that the more they listen to EM, the higher familiarity score they obtained.
In fact, participants who self-declared to listen to EM with a high frequency (4 and 5 in the 5-point Likert scale) on average obtained 0.34 on the familiarity score.
Ones with medium frequency got on average 0.19, while participants with low frequency 0.1, and the ones who almost no listen to EM obtained 0.05 (respectively self-declaring 3, 2, and 1 on the Likert scale).
Besides, we analyze the correlation between the familiarity score and the other self-declared music information finding a positive correlation with the variety of EM listened (Pearson’s $\rho=.40$, $p<.001$), the frequency of EM listening ($\rho=.39$, $p<.001$), the variety of music taste ($\rho=.22$, $p<.05$), and the frequency of music playing ($\rho=.19$, $p<.05$).
On the contrary, we found no evidence of a correlation between receiving formal music training and the familiarity score ($\rho=-.08$, $p=.42$).
These findings suggest that the familiarity score can be a further dimension to add when assessing participants' domain knowledge within EM, and it has the advantage of not being a self-declared attribute. 

\subsubsection{Participant Clustering and Demographic Grouping}
\label{clustering}
We performed an automatic clustering procedure to group participants based on their EM domain knowledge and familiarity.
To do that, we used the self-declared information about the frequency of listening to electronic music and taste variety, and the familiarity score.
We selected the k-medoids clustering technique (with k=3) \cite{park2009}, implemented in the $scikit\_extra$ Python package,\footnote{\url{https://scikit-learn-extra.readthedocs.io/en/latest/modules/cluster.html\#k-medoids}} because of its lower sensibility to outliers in comparison to other techniques such as K-means.
Additionally, we removed outliers identified as the points with a distance from the medoids greater than the average distance plus the standard deviation.
In this manner, we have been able to define three groups of participants respectively with low (group G1, size N=39), medium (G2, N=27), and high (G3, N=30) domain knowledge.
We additionally created four other groups based on demographic information for analyzing participants’ ratings.
First, we include in the WEIRD group (G4, N=84) participants coming from Europe and North America, holding a bachelor’s degree or a higher title.
The remaining participants were grouped in the non-WEIRD group (G5, n=24).
Then, we created two other groups dividing participants by age: age between 18 and 35 (G6, N=70), and age over 35 (G7, N=42).
While participants can belong \lorenzo{to} only one of the \lorenzo{groups} based on the domain knowledge (G1, G2, or G3), in the case of the demographic groups participants may have been included in \lorenzo{either G4 or G5}, and at the same time in \lorenzo{either G6 or G7}. 
This double-assignment is motivated by the purpose of analyzing different populations, divided by origin in group G4 and G5, and age in group G6 and G7, in which however a participant can coexist.
Participants who chose to not disclose such demographic information at the beginning of the survey were not included in these groupings.

\subsection{Agreement with Diversity Metrics}
\label{agreement}

\begin{table}[b]
  \caption{Percentage of users indicating which list was ``most diverse'' and pairwise distances $d_i$ of each of the two lists. Two aggregation methods are shown: $\operatorname{min}(d_i)$ is the minimum distance within a list, and $\operatorname{avg}(d_i)$ is the arithmetic average of the distances. For every list, we report the participants' agreement (\%agree). For every pair, larger distance values are in \textbf{bold}, while the list with greater participant agreement is in \textit{italic}. For a detailed description of the tasks see Section \ref{survey}.}.
  \label{tab:distances}
  \begin{tabular}{llcccccc}
\toprule
&Chosen&&$min(d_i)$&&&$avg(d_i)$& \\
Task&list ($\%agree$)&$cosine_{HL}$&$cosine_{LL}$&$Goodall$&$cosine_{HL}$&$cosine_{LL}$&$Goodall$\\
\midrule
TRACK1&A (7\%)&.03&.03&-&.08&.05&-\\
&\textit{B (89\%)}&\textbf{.12}&\textbf{.10}&-&\textbf{.30}&\textbf{.19}&-\\
TRACK2&\textit{A (85\%)}&\textbf{.16}&\textbf{.07}&-&\textbf{.44}&\textbf{.27}&-\\
&B (10\%)&.05&.07&-&.14&.16&-\\
TRACK3&\textit{A (76\%)}&.03&\textbf{.15}&-&\textbf{.26}&\textbf{.28}&-\\
&B (16\%)&\textbf{.09}&.10&-&.17&.22&-\\
TRACK4&A (30\%)&\textbf{.12}&.05&-&.20&.20&-\\
&\textit{B (58\%)}&.10&\textbf{.14}&-&\textbf{.29}&\textbf{.23}&-\\
\midrule
ARTIST1&A (3\%)&-&-&.41&-&-&.60\\
&\textit{B (94\%)}&-&-&\textbf{.79}&-&-&\textbf{.91}\\
ARTIST2&A (9\%)&-&-&.41&-&-&.60\\
&\textit{B (86\%)}&-&-&\textbf{.63}&-&-&\textbf{.79}\\
ARTIST3&\textit{A (70\%)}&-&-&\textbf{.66}&-&-&\textbf{.76}\\
&B (20\%)&-&-&.32&-&-&.72\\
ARTIST4&A (15\%)&-&-&\textbf{.67}&-&-&.81\\
&\textit{B (78\%)}&-&-&.66&-&-&\textbf{.88}\\
\midrule
COMBO1&A (19\%)&.04&.08&.54&.13&.11&.66\\
&\textit{B (69\%)}&\textbf{.09}&\textbf{.19}&\textbf{.61}&\textbf{.28}&\textbf{.30}&\textbf{.86}\\
COMBO2&A (18\%)&\textbf{.11}&.06&.40&\textbf{.39}&.18&.60\\
&\textit{B (72\%)}&.02&\textbf{.14}&\textbf{.42}&.25&\textbf{.24}&.60\\
COMBO3&A (43\%)&\textbf{.09}&.08&\textbf{.58}&.16&.17&\textbf{.79}\\
&\textit{B (50\%)}&.01&\textbf{.17}&.26&\textbf{.27}&\textbf{.21}&.71\\
COMBO4&\textit{A (55\%)}&\textbf{.13}&\textbf{.14}&.35&\textbf{.42}&\textbf{.35}&.54\\
&B (33\%)&.09&.04&\textbf{.65}&.19&.17&\textbf{.86}\\
    \bottomrule
  \end{tabular}
\end{table}

The results of the diversity estimated by means of the distance metrics are shown in Table \ref{tab:distances}, together with the participants' percentage agreement.
In addition, Table \ref{tab:mannwhitney} (see Appendix \ref{appendixA}) provides the results of the Mann-Whitney U test \cite{Mann1947}, to assess if differences between distances in the lists are statistically significant.

Regarding the agreement, we observe that in the \textit{TRACK} task the greater agreement is obtained where the two aggregation mechanisms ($\operatorname{min}(d_i)$ and $\operatorname{avg}(d_i)$, see Section \ref{distancemetrics}) agreed on which list was the more diverse.
Such finding is also confirmed by the statistically significant difference in the two first pairs of lists (\textit{TRACK1} and \textit{TRACK2}).
In contrast, looking at the other two pairs (\textit{TRACK3} and \textit{TRACK4}) we observe a decrease in the agreement in participants’ ratings where the aggregation mechanisms return contrasting results.
In addition, for such pairs the result of the Mann-Whitney U test does not suggest a significant difference between the lists presented either in terms of LL or HL features.
When asked for feedback on the selection strategy for this task, participants confirm that the following factors had a strong-medium influence on their choices: instrument and samples (83\% participants), sub-genre or sub-style (82\%), tempo/BPM (80\%),  feeling, emotion and mood (70\%).

For the \textit{ARTIST} tasks, we observe an analogous behaviour.
Indeed, the greater agreement between participants is found for the first two pairs where there is a large difference in the Goodall metric (\textit{ARTIST1} and \textit{ARTIST2}), while in the other two (\textit{ARTIST3} and \textit{ARTIST4}), the agreement decreases.
According to the participants’ feedback on their selection strategy, generational differences between artists had a strong-medium impact only for a few of them (13\%), while in contrast artists’ origin and nationality (82\%), gender (79\%), and color and skin tone (77\%) were considered key factors for diversifying the lists.

In the combined task (\textit{COMBO}), different factors seem to be interplaying.
In the first pair \textit{COMBO1} participants had a high agreement, and it is the only case where all the metrics and the aggregation mechanism agree on which list is more diverse.
\textit{COMBO2} represents an anomaly among the sets proposed. Indeed, both lists were formed by tracks similarly varied in terms of music features (no significant difference in terms of LL or HL features), but a list was formed by four tracks by white male EM artists versus a list formed by four tracks by white not-male artists.
As result, in \textit{COMBO2} lists were similarly different in terms of artists’ attributes, and according to the Goodall metric their difference was not significant.
However, for most of the participants (72\%) the list with not-male artists was perceived as being more diverse than the one with only male artists.
In the case of the \textit{COMBO3} and \textit{COMBO4} task, the low agreement corresponds to mixed results of the aggregation mechanisms, in line with the results found in the \textit{TRACK} and \textit{ARTIST} task.

When asked to indicate which among artist and track diversity they considered most in the last task, half of the participants indicated that they based their choices only or mostly on track diversity (50\%), few participants only or mostly on artist diversity (13\%), and the rest affirmed to consider both equally (37\%).

A limitation of the results proposed until now derives from the use as a yardstick of the percentage agreement among participants.
Indeed, such estimation does not take into account the agreement that would be expected by chance, hence the final results may be overestimated \cite{Hallgren2012}.
To cope with this issue, in the following section we present the obtained results using reliability analysis.

\subsection{Reliability Analysis}
\label{reliability}

We use Cohen's kappa \cite{Cohen1960} to estimate the pairwise agreement between a participant and what the most diverse list should be according to the metrics (metric reliability -- MR). We then report the average over each participant group, as described in Section \ref{clustering}.
Similarly, we measure inter-rater reliability (IRR) by computing the pairwise agreement between participants using Cohen's kappa, then averaged to provide a single value.
This value is equivalent to the Light’s kappa \cite{Hallgren2012}, and we report in Table \ref{tab:metricrater} and Table \ref{tab:interrater} such kappa for the MR and IRR respectively.
In addition to kappa, we also report the 95\% confidence interval obtained by bootstrapping \cite{Efron1992}.
The interpretation of kappa has been often debated, and there is no universally accepted scale \cite{Mchugh2012}.
For the sake of clarity, we refer to Cohen's suggested scale while interpreting the results obtained: values $\leq$ 0 indicate no agreement; .01–.20 none to slight; .21–.40 fair; .41–.60 moderate; .61–.80 substantial; .81–1.00 almost perfect agreement.

Results from the MR analysis of the \textit{TRACK} task indicate how much the participants’ perceptions agree with the diversity estimated with the cosine distance using the audio track features.
We observe three major findings: 1) participants with low domain knowledge (G1) agree more in comparison to participants with high domain knowledge (G3); 2) participants in the WEIRD group (G4) agree more in comparison with one in the non-WEIRD group (G5); 3) participants with age between 18-35 (G6) agree more than participants with age over 35 (G7).
Generally, the agreement in this task ranges from fair to substantial, almost perfect in very few cases.

\begin{table}[ht]
  \caption{Metric reliability (MR) results. The Light's kappas and 95\% confidence interval are reported separately for each group and each task. Values are in \textbf{bold}(\textit{italic}) when they are 0.05 higher(lower) than the kappa measured across all participants.
  "T" refers to the \textit{TRACK} task, "A" to the \textit{ARTIST} task, "C" to the \textit{COMBO} task.
  For details on the tasks see Section \ref{survey}.
  G1, G2, G3 are groups of users respectively with low, medium, and high domain knowledge; G4 is WEIRD and G5 non-WEIRD; G6 is $\le 35$ years old, G7 is $> 35$ years old.
  For details on the groups see Section \ref{clustering}.}
  \label{tab:metricrater}
  \begin{minipage}{\columnwidth}
  \begin{center}
  \begin{tabular}{lcccccccccc}
    \toprule
    &\textit{T}&\textit{T1-2}&\textit{T3-4}&\textit{A}&\textit{A1-2}&\textit{A3-4}&\textit{C}&\textit{C1-2}&\textit{C3-4}\\
    \midrule
    G1&.63$\pm$.02&.81$\pm$.02&.46$\pm$.03&.64$\pm$.02&.87$\pm$.02&.57$\pm$.03&.25$\pm$.03&.56$\pm$.02&\textit{.09}$\pm$ .04\\
    G2&.65$\pm$.02&\textit{.72}$\pm$.03&\textbf{.60}$\pm$.04&.57$\pm$.04&.85$\pm$.02&.59$\pm$.04&.26$\pm$.03&.56$\pm$.04&\textbf{.20}$\pm$.05\\
    G3&\textit{.48}$\pm$.03&.73$\pm$.03&\textit{.27}$\pm$.05&\textbf{.68}$\pm$.03&.83$\pm$.02&\textbf{.66}$\pm$.03&.20$\pm$.03&.50$\pm$.03&.17$\pm$.03\\
    \midrule
    G4&.61$\pm$.01&.80$\pm$.01&.44$\pm$.02&.63$\pm$.01&.83$\pm$.01&.61$\pm$.01&.22$\pm$.01&.54$\pm$.01&\textit{.10}$\pm$.01\\
    G5&\textit{.51}$\pm$.03&\textit{.69}$\pm$.04&\textit{.33}$\pm$.05&\textit{.53}$\pm$.04&.83$\pm$.03&\textit{.47}$\pm$.05&.28$\pm$.04&\textit{.42}$\pm$.03&\textbf{.26}$\pm$.05\\
    \midrule
    G6&.63$\pm$.01&.81$\pm$.01&.47$\pm$.02&.61$\pm$.01&.86$\pm$.01&.55$\pm$.02&.25$\pm$.01&.54$\pm$.02&.17$\pm$.02\\
    G7&.55$\pm$.02&.74$\pm$.02&\textit{.38}$\pm$.02&.59$\pm$.02&.81$\pm$.02&.60$\pm$.03&.22$\pm$.02&.50$\pm$.02&.10$\pm$.03\\
    \midrule
    \textit{all}&.59$\pm$.01&.77$\pm$.01&.43$\pm$.01&.60$\pm$.01&.83$\pm$.01&.56$\pm$.01&.24$\pm$.01&.52$\pm$.01&.15$\pm$ .01\\
\bottomrule
  \end{tabular}
  \end{center}
  \end{minipage}
\end{table}

\begin{table}[ht]
  \caption{Inter-rater reliability (IRR) results. The Light's kappas and 95\% confidence interval are reported separately for each group and each task. Kappas are in \textbf{bold}(\textit{italic}) when higher(lower) than 0.05 as to the kappa measured across all participants. "T" refers to the \textit{TRACK} task, "A" to the \textit{ARTIST} task, "C" to the \textit{COMBO} task. For a detailed description of the groups see Section \ref{clustering}, for the tasks see Section \ref{survey}.}
  \label{tab:interrater}
  \begin{minipage}{\columnwidth}
  \begin{center}
  \begin{tabular}{lcccccccccc}
    \toprule
    &\textit{T}&\textit{T1-2}&\textit{T3-4}&\textit{A}&\textit{A1-2}&\textit{A3-4}&\textit{C}&\textit{C1-2}&\textit{C3-4}\\
    \midrule
    G1&\textbf{.45}$\pm$.02&\textbf{.66}$\pm$.03&.27$\pm$.03&.41$\pm$.03&\textbf{.76}$\pm$.03&.34$\pm$.03&.09$\pm$.01&.30$\pm$.03&\textit{.03}$\pm$.01\\
    G2&\textbf{.44}$\pm$.03&\textit{.53}$\pm$.04&\textbf{.40}$\pm$.03&.37$\pm$.03&.72$\pm$.04&\textbf{.45}$\pm$.03&.14$\pm$.02&\textbf{.33}$\pm$.04&.09$\pm$.02\\
    G3&\textit{.30}$\pm$.02&\textit{.55}$\pm$.03&\textit{.10}$\pm$.03&\textbf{.46}$\pm$.03&.69$\pm$.04&\textbf{.43}$\pm$.04&.06$\pm$.02&.24$\pm$.03&\textbf{.17}$\pm$.02\\
    \midrule
    G4&.41$\pm$.01&.64$\pm$.01&.23$\pm$.01&.41$\pm$.01&.70$\pm$.01&\textbf{.39}$\pm$.01&.10$\pm$.01&.28$\pm$.01&.09$\pm$.01\\
    G5&\textit{.32}$\pm$.03&\textit{.50}$\pm$.04&.20$\pm$.04&\textit{.30}$\pm$.04&.69$\pm$.05&.27$\pm$.04&.09$\pm$.02&\textit{.17}$\pm$.03&\textbf{.15}$\pm$.03\\
    \midrule
    G6&\textbf{.44}$\pm$.01&\textbf{.66}$\pm$.01&.26$\pm$.01&.38$\pm$.02&.73$\pm$.02&.33$\pm$.02&.11$\pm$.01&.30$\pm$ .02&.12$\pm$.01\\
    G7&\textit{.33}$\pm$.02&\textit{.56}$\pm$.03&.19$\pm$.02&.38$\pm$.02&.67$\pm$.03&\textbf{.39}$\pm$.02&.09$\pm$.01&.24$\pm$.02&.06$\pm$.01\\
    \midrule
    \textit{all}&.39$\pm$.01&.61$\pm$.01&.23$\pm$.01&.38$\pm$.01&.70$\pm$.01&.34$\pm$.01&.10$\pm$.00&.27$\pm$.01&.10$\pm$.01\\
    \bottomrule
  \end{tabular}
  \end{center}
  \end{minipage}
\end{table}

For the \textit{ARTIST} task, we observe a different behaviour than the former task when comparing the perceived diversity with the diversity estimated by means of the Goodall metric.
Indeed, in this case the domain knowledge is not so influential for determining the artist diversity, being not the task ratings based on any music-related aspect.
Still, WEIRD and younger participants seem to agree more in comparison to their complementary groups.
For this task, we observe moderate to almost perfect agreement for every analyzed group.

Regarding the \textit{COMBO} task, we observe an overall decrease of agreement level, expected as the complexity of this task increases in comparison to the former two tasks.
Only in the case of a generational difference (G6 versus G7), we continue to see a greater agreement for the youngest participants.
Looking at all the tasks, the kappa values for the pairs designed for having a more substantial difference (1-2) are greater than for the pairs where the difference between lists was slighter (3-4), confirming the task selection strategy presented in Section \ref{taskselection}.
Lastly, we notice that results from G4 (WEIRD) and G6 ($\le$ 35 years old) groups are almost in line with the results for all participants, being the majority of the subjects of our study belonging to such demographic groups.

The IRR analysis gives us a similar picture of the MR analysis, as shown in Table \ref{tab:interrater}.
Indeed, the agreement in the \textit{TRACK} task is higher for participants with low domain knowledge, belonging to the WEIRD group, and aged between 18 and 35.
Again, such difference is not present in the \textit{ARTIST} task when comparing groups with different domain knowledge, while it is found comparing groups with different demographics.
The inter-rater agreement in the \textit{COMBO} task decreases, as observed in the MR analysis, and also the difference between pairs 1-2 and 3-4 is confirmed for the three tasks.
In general, the inter-rater agreement can be interpreted as fair to moderate in the \textit{TRACK} and \textit{ARTIST} task, with few exceptions where it is substantial. For the \textit{COMBO} task, in most of the cases there is none to slight inter-rater agreement.

\section{Discussion and Limitations}
\label{discussion}

In this section, we discuss our observations and try to answer the research questions about how different elements may influence perceived diversity.
We also describe several limitations of this study.

With respect to RQ1 --- \textit{To what extent tracks’ audio features and artists’ attributes can be used to assess the perceived diversity?} --- we have shown that audio-based features have some applicability for determining diversity, particularly when the case is clear enough that most people would agree that one list is more diverse than another.
For this task, HL features may have an advantage over LL features in terms of interpretability and semantic meaning.
However, LL features may be more discriminatory in assessing the diversity of a list, being perceptually less prone to subjective assessment in comparison to HL (e.g. differences in BPM are less subjective than differences in the happiness of a song), and being the HL constructs conceptually problematic by themselves (e.g. the definition of what a ``happy song'' is) \cite{Liem2020}.
Artists’ attributes can be used to model computationally a degree of diversity, which results from the reliability analysis confirmed to be shared among people from the same social group.
When mixing different diversity dimensions (from tracks and artists), the increased complexity of the task results in a decreased agreement. 

With respect to RQ2 --- \textit{To what extent domain knowledge and familiarity influence participants’ perceptions of diversity?} --- domain knowledge can be construed in multiple ways, including familiarity with artists or tracks.
These are somewhat correlated, and all of them influence perceptions of diversity.
For instance, self-declared music taste diversity within a music genre (electronic music) is correlated with familiarity and interacts with the perceptions of diversity.
Moreover, track diversity perceptions strongly depend on domain knowledge and familiarity. 
Indeed, listeners with low domain knowledge are more easily satisfied with the diversity of a set of tracks and tend to agree more among them, while listeners with high domain knowledge are less easily satisfied in this regard and tend to agree less among them.
Whilst these results might be counterintuitive at first, they agree with observations from previous work where it has been shown that experts tend to agree less on other types of music-related assessment, such as the emotion perceived while listening to music \cite{Schedl2018}.
On the contrary, domain knowledge seems to have a limited effect on artist diversity compared to its effects over track diversity.

\textbf{Limitations} of the presented study are several and discussed hereafter.
The generalization of the findings is limited by at least two factors.
First, the participant population is strongly biased towards WEIRD societies, and consequently, results cannot provide a global validity.
At the very least, a wider sample of participants coming from non-Western countries would be necessary to generalize our findings.
Second, the choice to focus on electronic music has been motivated by the goal of assessing the relationship between domain knowledge and perceived diversity in a controlled environment.
However, to validate our hypothesis, experimentation with other music genres should be foreseen.

A limitation in the design of the experiment is to explicitly ask participants to not consider popularity, likeness, or familiarity with the tracks and artists while assessing the diversity of the lists. 
Indeed, those aspects may interplay with the perception of diversity, and it may be hard to exclude them during the assessment.
Further investigation is required to understand their role and relationship with diversity.
In addition, listeners’ characterization can be extended considering other music-related attributes, such as musical sophistication \cite{Mullensiefen2014}, but also considering other types of signals, such as personal listening histories to infer music tastes.
Being the participation in our study voluntarily, it has been designed to limit the time needed to complete the survey to 30-35 minutes.
This also determined the small number of sets for which participants have been asked to provide their opinion. Recruiting participants by providing an economic reward can be a method to overcome such issues.
Lastly, the design of the study, the task and data selection, as well as computational aspects such as the features and metric selection cannot be immune to the authors’ standpoint.
We aim at improving these methods by confronting our approach with research communities from different fields.

\section{Conclusion and Future Work}
\label{conclusion}

Listeners, artists, and tracks, if imagined as \textit{closed doors} may limit the understanding of cultural networks, in which a dual nature is embedded, where shared interests constitute social groups, but also shared publics form genres and styles \cite{dimaggio2011}.
Differences in these networks constitute what diversity is, and even if historically such differences have been often used for creating boundaries among cultures, for instance between Western music and others, the same differences can serve as a tool to promote and support multi- and intra-cultural environments \cite{Grenier1989,Born2000,Serra2011, Clarke2015}.

Through a user study, we analyzed how different diversity dimensions interact and how they are perceived, to prove that diversity cannot be understood if not considering its multifaceted nature.
Even if features that are computable from the audio signal may help categorize music lists into coarse categories by diversity, complexities may arise while considering fine-grained ones.
Consequently, such features may help generate diversity-aware recommendations to users with less domain knowledge but will be less useful for more specialized users.
We showed that domain knowledge can be estimated by mixing self-declared answers and automatic scores, representing different aspects of familiarity and expertise within EM, and we provided a method for its assessment.

From another perspective, artists embed their diversity in their works and taking into account not only music features but also socially relevant characteristics can lead to new insights for designing diversity-aware music recommendations.
The scenario described in this experiment represents a particular case of multistakeholder environment \cite{Abdollahpouri2020}, where listeners represent the \textit{consumers} of the content created by the artists, the \textit{providers} in our settings. 
RS research in this area is recently emerging as a new sub-field wherein all the parts interacting are considered in the whole process of design, development, and evaluation of RS. 
Under this lens, our study of the characteristics of the several actors involved in Music RS may offer a new perspective for multistakeholder recommender systems diversity-related evaluation.

In order to cope with the limitations in terms of generalizability of the findings presented in this work, we plan to replicate the user study both considering a music genre different from electronic music and selecting a different participant population, not biased towards Western countries.

\noindent{\textbf{Reproducibility.}}
To encourage reproducibility of the results, the code and data used in our study are publicly available.\footnote{\lorenzo{\url{https://github.com/LPorcaro/music-diversity-analysis}}}

\begin{acks}

\lorenzo{This work is partially supported by the European Commission under the TROMPA project (H2020 - grant agreement No.\ 770376).}

\lorenzo{This work is also partially supported by the HUMAINT programme (Human Behaviour and Machine Intelligence), Joint Research Centre, European Commission.}

\lorenzo{The project leading to these results received funding from "la Caixa" Foundation (ID 100010434), under the agreement LCF/PR/PR16/51110009.}
\end{acks}

\bibliographystyle{ACM-Reference-Format}
\bibliography{main}

\newpage
\appendix
\section{Supplementary Material}
\label{appendixA}
\begin{figure}[h]
\centering
\includegraphics[width=1\textwidth]{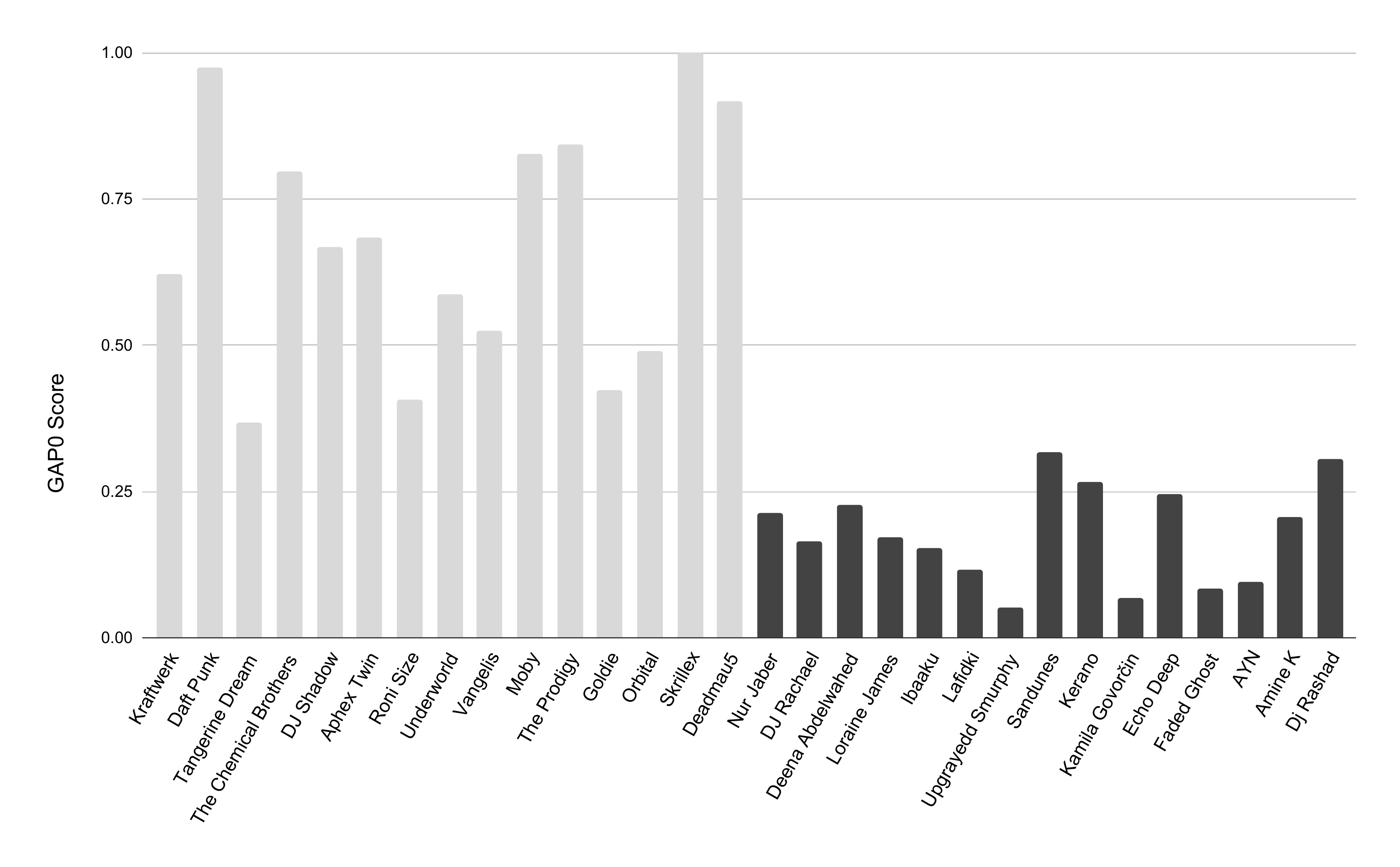}
\Description{List of artists and corresponding GAP0 score for assessing the participants' familiarity with EM artists. Artists selected within Allmusic most representative list are in light grey (left), while artists selected within the ones shown in the survey are in dark grey (right). “Skrillex” results to be the most popular, whilst “Upgrayedd Smurphy” the one with less social presence.}
\caption{List of artists and corresponding GAP0 score for assessing the participants' familiarity with EM artists. Artists selected within Allmusic most representative are in light grey (\textit{left}), while artists selected within the ones shown in the survey are in dark grey (\textit{right}). “Skrillex” results to be the most popular, whilst “Upgrayedd Smurphy” the one with less social presence.}
\label{fig:fig3}
\end{figure}

\begin{table}
  \caption{List of tracks included in the \textit{TRACK} task. (* Following the controversy about the previous alias used by the artist ("The Black Madonna") we report here the new alias. Such event occurs during the time when authors collected participants answers. For more details: \url{https://pitchfork.com/news/the-black-madonna-changes-name-to-the-blessed-madonna/}).}
  \label{tab:tracktask}
  \begin{tabular}{ccll}
    \toprule
    Task&List& Artist Name & Track Name\\
    \midrule
    TRACK1&List A&Resensed \& Asterz&Last Time\\
    &&KELTEK \& B-Front&Step Into The Game\\
    &&Hard Driver x Max Enforcer&Before I Wake\\
    &&Exilium &I Got All The Time\\
    &List B&999999999&X0004000X\\
    &&HOSH \& 1979 Feat. Jalja &Midnight (The Hanging Tree)\\
    &&Tibi Dabo &Komorebi\\
    &&Kid Francescoli &Moon\\
    \midrule
    TRACK2&List A&Pretty Light&Yellow Bird\\
    &&Bonobo&Kong\\
    &&Technoboy&Ti sento\\
    &&Laurent Garnier&The man with the red face\\
    &List B&Amelie Lens&The Future\\
    &&Nicole Moudaber&Own\\
    &&Russell E.L. Butler&Gulf Stream\\
    &&Satoshi Tomiie&Resonant\\
     \midrule
    TRACK3&List A&The Blessed Madonna*&Stay\\
    &&Beatrice Dillon&Workaround One\\
    &&Miss DJAX&Headbangin' \\
    &&Helena Hauff&Spur\\
    &List B&Nur Jaber&Until I Collapse\\
    &&DJ Rachael&Silaye\\
    &&Deena Abdelwahed&Ababab\\
    &&Nina Kraviz&Desire\\
     \midrule
    TRACK4&List A&Echo Deep&Maasai Groove\\
    &&Amine K&Dar Gnawa\\
    &&Ricardo Villalobos&Lazer@Present\\
    &&Solomun&The way back\\
    &List B&Aphex Twin&Windowlicker\\
    &&Dubfire&Roadkill\\
    &&Orieta Chrem&Delfines\\
    &&Dj Rashad&Feelin\\
    \bottomrule
  \end{tabular}
\end{table}

\begin{table}
  \caption{List of artists included in the \textit{ARTIST} task. Artists' gender is reported as  male (M), female (F), or non-binary (X). Skin tone is categorized following the Fitzpatrick scale \cite{Fitzpatrick1988}. Debut refers to the decade in which the artists started playing.}
  \label{tab:artisttask}
  \begin{tabular}{rlcccccc}
    \toprule
    Task / List&Artist Name&Birthplace&Gender&\makecell{Skin \\Tone}&Debut\\
    \midrule
    ARTIST1 / List A&Pretty Lights&United States&M&I&2000'\\
    &Bonobo&United Kingdom&M&II&1990'\\
    &Technoboy&Italy&M&II&1990'\\
    &Laurent Garnier&France&M&II&1980'\\
    List B&Amelie Lens&Belgium&F&II&2010'\\
    &Nicole Moudaber&Nigeria&F&III&2000'\\
    &Russell E.L. Butler&Bermuda&X&IV&2010'\\
    &Satoshi Tomiie&Japan&M&II&1980'\\
    \midrule
    ARTIST2 / List A&The Blessed Madonna*&United States&X&I&2010'\\
    &Beatrice Dillon&United Kingdom&F&I&2010'\\
    &Miss DJAX&The Netherlands&F&II&1980'\\
    &Helena Hauff&Germany&F&II&2010'\\
    List B&Nur Jaber&Lebanon&F&II&2010'\\
    &DJ Rachael&Uganda&F&VI&1990'\\
    &Deena Abdelwahed&Tunisia&X&III&2010'\\
    &Nina Kraviz&Russia&F&I&2000'\\
    \midrule
    ARTIST3 / List A&Loraine James&United Kingdom&X&V&2010'\\
    &Ibaaku&Senegal&M&VI&2010'\\
    &Lafidki&Cambodia&M&III&2010'\\
    &Upgrayedd Smurphy&Mexico&F&I&2010'\\
    List B&Sandunes&India&F&III&2010'\\
    &Kerano&India&M&III&2010'\\
    &Kamila Govorčin&Chile&F&III&2010'\\
    &Nicolas Jaar&United States&M&II&2010'\\
    \midrule
    ARTIST4 / List A&Echo Deep&South Africa&M&V&2010'\\
    &Amine K&Morocco&M&IV&2000'\\
    &Ricardo Villalobos&Chile&M&III&1990'\\
    &Solomun&Bosnia-Herzegovina&M&II&2000'\\
    List B&Aphex Twin&Ireland&M&I&1980'\\
    &Dubfire&Iran&M&III&1990'\\
    &Orieta Chrem&Peru&F&II&2000'\\
    &Dj Rashad&United States&M&VI&1990'\\
    \bottomrule
  \end{tabular}
\end{table}

\begin{table}
  \caption{List of tracks and artists included in the \textit{COMBO} task.}
  \label{tab:bothtask}
  \begin{tabular}{ccll}
    \toprule
    Task&List& Artist Name & Track Name\\
    \midrule
    COMBO1&List A&Skrillex&Bangarang\\
    &&Haezer&Bass Addict\\
    &&Clean Tears&Bouquet\\
    &&Carl Nunes&Limitless\\
    &List B&Le Bask&Hardchoriste\\
    &&Faded Ghost&Moon Rain\\
    &&Andres Gil&Scene\\
    &&AYN&Where It Went\\
    \midrule
    COMBO2&List A&Pretty Light&Yellow Bird\\
    &&Bonobo&Kong\\
    &&Technoboy&Ti sento\\
    &&Laurent Garnier&The man with the red face\\
    &List B&The Blessed Madonna*&Stay\\
    &&Beatrice Dillon&Workaround One\\
    &&Miss DJAX&Headbangin' \\
    &&Helena Hauff&Spur\\
    \midrule
    COMBO3&List A&Nur Jaber&Until I Collapse\\
    &&DJ Rachael&Silaye\\
    &&Deena Abdelwahed&Ababab\\
    &&Nina Kraviz&Desire\\
    &List B&Sandunes&Exit Strategy\\
    &&Kamila Govorčin&Daydream\\
    &&Nicolas Jaar&A time for us\\
    &&Kerano&Thirsty\\
    \midrule
    COMBO4&List A&Ibaaku&Discuting Food\\
    &&DJ Rachael&Silaye\\
    &&Echo Deep&Maasai Groove\\
    &&Bosaina&Abalone on the Grass\\
    &List B&Amelie Lens&The Future\\
    &&Aphex Twin&Windowlicker\\
    &&Ricardo Villalobos&Lazer@Present\\
    &&Peggy Gou&Starry Night\\
    \bottomrule
  \end{tabular}
\end{table}

\begin{table}
  \caption{List of features used for computing the distance using the cosine metric.}
  \label{tab:features}
  \begin{tabular}{cll}
    \toprule
    High-Level&danceability&mood sad\\
    &mood aggressive&timbre\\
    &mood happy&tonal / atonal\\
    &mood party&voice / instrumental\\
    &mood relaxed&\\
    \midrule
    Low-Level&average loudness&spectral energy\\
    &barkbands crest&spectral energyband high\\
    &barkbands flatness db&spectral energyband low\\
    &barkbands kurtosis&spectral energyband middle high\\
    &barkbands skewness&spectral energyband middle low\\
    &barkbands spread&spectral entropy\\
    &dissonance&spectral flux\\
    &dynamic complexity&spectral kurtosis\\
    &erbbands crest&spectral rms\\
    &erbbands flatness db&spectral rolloff\\
    &erbbands kurtosis&spectral skewness\\
    &erbbands skewness&spectral strongpeak\\
    &erbbands spread&zero crossing rate\\
    &hfc&chords changes rate\\
    &melbands crest&chords number rate\\
    &melbands flatness db&chords strength\\
    &melbands kurtosis&hpcp crest\\
    &melbands skewness&hpcp entropy\\
    &melbands spread&tuning diatonic strength\\
    &pitch salience&tuning equal tempered deviation\\
    &silence rate 30dB&tuning frequency\\
    &silence rate 60dB&tuning nontempered energy ratio\\
    &spectral centroid&onset rate\\
    &spectral decrease&BPM\\
    \bottomrule
  \end{tabular}
\end{table}

\begin{table}
  \caption{Report of Mann–Whitney U test. The results for every comparison presented to participants are shown, separately for the cosine metric computed using HL ($cosine_{HL}$) and LL ($cosine_{LL}$) features, and the Goodall metric computed with the artists' attributes. Together with the median distance for every list, the U-statistics, the significance at $p=0.05$ and $p=0.1$, and the effect size (computed using the Wendth formula \cite{Wendt1972}) are reported. For every pair, the list with the highest median is in \textbf{bold}.}
  \label{tab:mannwhitney}
  \begin{tabular}{llcccccc}
    \toprule
    &Task&\makecell{Median($d_i$)\\List A}&\makecell{Median($d_i$)\\List B}&U &\makecell{Significance \\at p=0.05}&\makecell{Significance \\at p=0.1}&Effect Size\\
    \midrule
    $cosine_{HL}$&TRACK1&.08&\textbf{.30}&1&True&True&.94\\
    &TRACK2&\textbf{.50}&.14&3&True&True&.83\\
    &TRACK3&\textbf{.28}&.13&10&False&False&.44\\
    &TRACK4&.17&\textbf{.33}&9&False&False&.5\\
    &COMBO1&.13&\textbf{.28}&8&False&False&.56\\
    &COMBO2&\textbf{.44}&.28&9&False&False&.5\\
    &COMBO3&.13&\textbf{.27}&12&False&False&.33\\
    &COMBO4&\textbf{.43}&.19&6&False&True&.67\\
    \midrule
    $cosine_{LL}$&TRACK1&.05&\textbf{.18}&0&True&True&1\\
    &TRACK2&\textbf{.31}&.16&5&True&True&.72\\
    &TRACK3&\textbf{.27}&.20&12&False&False&.33\\
    &TRACK4&\textbf{.23}&.22&15&False&False&.17\\
    &COMBO1&.10&\textbf{.32}&0&True&True&1\\
    &COMBO2&.20&\textbf{.21}&13&False&False&.27\\
    &COMBO3&.16&\textbf{.20}&10&False&False&.44\\
    &COMBO4&\textbf{.38}&.17&4&True&True&.78\\
    \midrule
    \textit{Goodall}&ARTIST1&.68&\textbf{.90}&0&True&True&1\\
    &ARTIST2&.60&\textbf{.82}&5&False&True&.72\\
    &ARTIST3&\textbf{.80}&.79&17&False&False&.03\\
    &ARTIST4&.82&\textbf{.91}&11&False&False&.39\\
    &COMBO1&.83&\textbf{.89}&9&False&False&.5\\
    &COMBO2&\textbf{.63}&.60&15&False&False&.17\\
    &COMBO3&\textbf{.83}&.78&17&False&False&.05\\
    &COMBO4&.57&\textbf{.88}&0&True&True&1\\
    \bottomrule
  \end{tabular}
\end{table}

\end{document}